%% file: procn.tex
\documentclass{PoS}

\title{Relativistic, model-independent, three-particle quantization condition}

\ShortTitle{Three-particle quantization condition}

\author{Maxwell T. Hansen\\
        Physics Department, University of Washington, Seattle, WA 98195-1560, USA\\
        E-mail: \email{mth28@uw.edu}}

\author{Stephen R. Sharpe\\
        Physics Department, University of Washington, Seattle, WA 98195-1560, USA\\
        E-mail: \email{srsharpe@uw.edu}}

\usepackage[sort&compress,numbers]{natbib}
\usepackage{mathtools}
\usepackage{amsfonts} 
\usepackage{amssymb} 
\usepackage{amsmath} 
\usepackage{graphicx} 
\usepackage{latexsym} 
\usepackage{verbatim} 

\newcommand{\Kdf}[1]{  \widetilde{\mathcal K}_{\mathrm{df}, 3 \rightarrow 3 #1}}

\abstract{This is a combined write-up for two talks which were given
  consecutively and which described different aspects of the same
  topic. We present a generalization of L\"uscher's relation between
  the finite-volume spectrum and \(S\)-matrix to three particles. Specifically, we consider a scalar field theory, which
  has a \(\mathbb{Z}_2\) symmetry that prevents even/odd coupling. The theory is assumed to have no two-particle bound states and to have two-particle phase shifts that are bounded by $\pi/2$ in the regime of elastic scattering. Considering center of mass energies between one
  and five particle masses, we evaluate a three-to-three finite-volume
  correlator to all orders in perturbation theory. Only terms which
  are exponentially suppressed in volume are neglected. From poles in
  the correlator we then determine the relation between finite-volume
  spectrum and scattering quantities. In this analysis one must carefully treat
  the unitary cusp at two-particle threshold. This point, which was
  neglected in the conference talks, is described in some detail
  here. We also describe an important check on our main result by
  reproducing the large volume expansion of the energy shift from the
  three-particle threshold. This is found to be consistent with previous work
  through four non-trivial orders. 
}

\FullConference{The 31st International Symposium on Lattice Field Theory\\
                 July 29 - August 3,  2013\\
                 Mainz, Germany}

\begin{document}

\section{Introduction}

Numerical lattice QCD (LQCD) is the only known systematic method for
determining non-perturbative, low-energy properties of the strong
interaction. However, it is only possible to determine QCD correlators
defined in Euclidean time. This presents a serious challenge for
extracting multi-hadron scattering amplitudes. 

In the case of two-to-two scattering, the issue was resolved by Martin
L\"uscher in a series of papers from 1986-1991
\cite{Luescher:1986n1,Luescher:1986n2,Luescher:1991n1}. His central
insight was that one can use finite spatial volume
as a tool to recover infinite-volume scattering
information. Specifically, he derived a relation between the
discrete finite-volume spectrum and the elastic scattering amplitude,
valid below the inelastic threshold.
This approach has since been implemented
in numerous numerical studies, allowing extraction of scattering
phases and determination of resonance properties.

Given the robust body of theoretical and numerical work in the
two-particle sector, it is natural to investigate whether the relation
between the finite-volume spectrum and scattering amplitudes can be
generalized to accommodate three (and higher) particle states. The
last few years have seen important developments in this 
direction~\cite{Polejaeva:2012,Briceno:2012rv}. However, a
complete method for extracting three-particle scattering
amplitudes is still unavailable. 

This implies, for example, that LQCD cannot yet offer predictions on
the mass and width of the \(\omega\) resonance, since it decays
predominantly into three pions. Similarly the Roper resonance, \(N(1440)\),
which decays with a \(40\%\) branching fraction into three-particle
states, cannot be rigorously investigated. The latter case is
especially interesting since its position below \(N(1535)\)
contradicts quark-model predictions, with more complicated
explanations so far reaching no consensus. 
Along similar lines, three body weak decays such as $K\to3\pi$ cannot
yet be investigated using LQCD.

Further motivation comes from the realization
that three-body forces are important for
understanding strongly interacting quantum systems such
as nuclei and neutron stars. Any effort to describe such
systems from first-principles QCD thus requires a method for extracting
three-particle scattering from lattice simulations. 
As a final motivation, we recall that
elastic phase shifts extracted from LQCD are currently limited to
energies below inelastic threshold. Only by generalizing the formalism
to include three or more particle-states can this range of
validity be extended.

In this work we take a step in this direction by deriving
a relation between the finite-volume spectrum of three particle
states and infinite-volume two-to-two and
three-to-three scattering, in the context of
a relativistic scalar field theory.
Here we can only sketch the derivation; details will be
given in a forthcoming article~\cite{inprogress}.
We also note that some technical aspects of the result have
been corrected since the talks, particularly those in 
section~\ref{sec:cusp}.

\section{Set-up}

We assume throughout a finite, cubic spatial volume with extent \(L\)
and with periodic boundary conditions. We demand that \(L\) is large
enough to neglect exponentially suppressed corrections of the form
\(e^{- mL}\), where \(m\) is the particle mass. Although the main
target of this formalism is finite-volume lattice calculations, we
assume here that discretization errors are small and have been
controlled elsewhere. We therefore work throughout in continuum field
theory (zero lattice spacing). We also differ from the standard
simulation set-up by working in Minkowski rather than Euclidean time,
with the time coordinate of infinite extent. Minkowski time turns out
to be convenient for our analysis and the distinction is irrelevant to
the final result.

We consider a scalar field theory describing particles of
mass \(m\). Thus, all results in this work are valid for identical
particles. The Lagrangian of the theory is arbitrary except that it
is invariant under a \(\mathbb Z_2\) symmetry that prevents coupling
between even- and odd-number particle states. (For the pion in QCD
this is G-parity.) 
Aside from the \(\mathbb Z_2\) symmetry, 
we make no restrictions on the Lagrangian. In
particular we include all vertices with an even number of scalar
fields and make no assumptions about relative coupling strengths.
We further require that the theory has no two-particle bound states and that the two-particle phase shifts do not have 
magnitude exceeding \(\pi/2\) for energies between the two-particle and the inelastic (four-particle)
thresholds. These criteria are necessary to prevent poles in the two-particle K-matrix, 
which would invalidate our present derivation.


Following \cite{Kim:2005} we determine the spectrum
from a finite-volume, Minkowski-time correlator
\begin{equation}
C_L(E, \vec P) \equiv \int_{L} d^4 x e^{i(-\vec P \cdot \vec x + E
  x^0)} \langle 0 \vert \mathrm{T} \sigma(x) \sigma^\dagger(0) \vert 0
\rangle \,.
\end{equation}
Here T indicates time-ordering and \(\sigma(x)\) is an interpolating
field, with spatial periodicity \(L\), that couples to states with an
odd number of scalar particles. The Fourier transform
restricts the states to have
total energy and momentum \((E, \vec P=2 \pi \vec n_P/L)\),
with $\vec n_P$ a vector of integers.
We denote by \(E^*\) the total energy in the center of mass
(CM) frame, (\(E^{*2}=E^2 - \vec P^2\)). 
The subsequent derivation holds only when we limit the total energy
to lie in the range \(m<E^*<5m\). This restricts the possible on-shell
intermediate states to those with three particles.

At fixed \(\{L, \vec P\}\), the spectrum of our theory is given by the
set of CM energies \(E^*_1, E^*_2, \cdots\) for which \(C_L \big
([E^{*2} + \vec P^2]^{1/2}, \vec P \big)\) has a pole. Thus our goal
is to determine the poles in $C_L$ when we include
all finite-volume contributions scaling as a power of \(1/L\).
The finiteness of the volume enters
our calculation through the prescription of summing (rather than
integrating) the spatial components of all loop momenta:
\begin{equation}
\int \frac{d^4k}{(2\pi)^4} \longrightarrow
 \frac{1}{L^3} \sum_{\vec k = 2 \pi \vec n/L} \int \frac{d k^0}{2 \pi}
 \mathrm{\ \ \ with\ \ \ } \vec n \in \mathbb Z^3 \,.
\end{equation}
We stress that when $L\to\infty$ the correlator does not have poles,
in our energy range, but only cuts.

\begin{figure}
\begin{center}
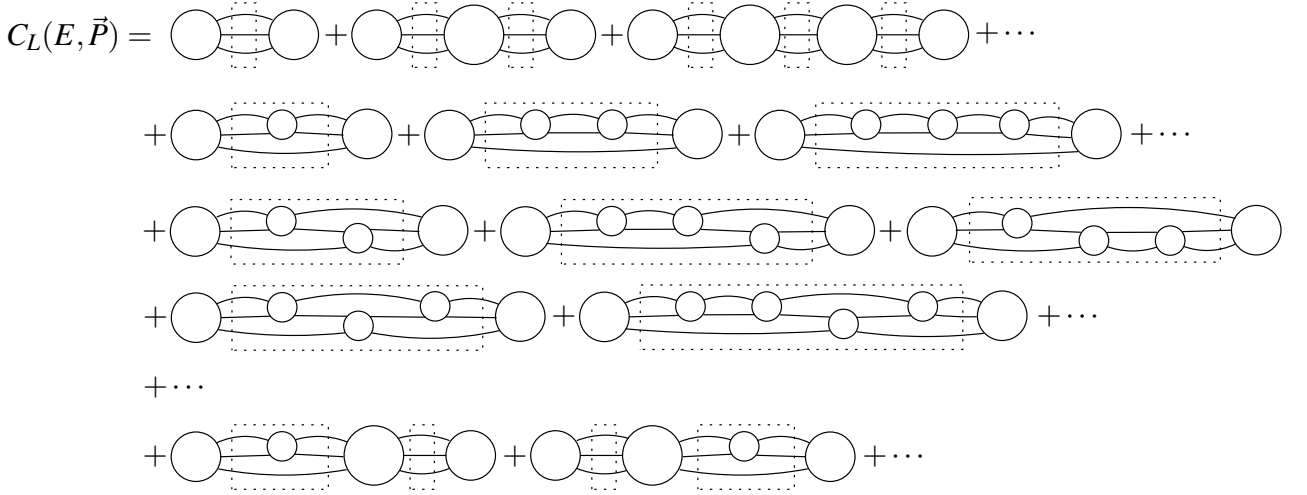
\caption{Skeleton expansion defining the finite-volume correlator. The
  leftmost circle in all diagrams represents the function 
  \(\widetilde \sigma\), while the rightmost represents
  \(\widetilde \sigma^\dagger\). Any insertion between these
  with four (six) legs represents a two-to-two (three-to-three)
  Bethe-Salpeter kernel  \(i B_{2 \rightarrow 2}\)
  (\(i B_{3 \rightarrow 3}\)). All lines represent
  fully-dressed propagators. Finally, dashed rectangles indicate
  that all loop momenta on enclosed propagators
  are summed rather than integrated.
  See text for further details.
}
\label{fig:fullskelexpansion}
\end{center}
\end{figure}

A key technical observation underlying our analysis is
that finite-volume momentum sums can be replaced by integrals
if the integrand is non-singular and smooth~\cite{Luescher:1986n2}.
This replacement leads
only to errors which are exponentially suppressed, and which we assume are
negligible. This motivates organizing the sum of diagrams contributing
to $C_L$ into the skeleton expansion shown in 
Fig.~\ref{fig:fullskelexpansion}.
Here we keep explicit all intermediate states which can go on shell,
while collecting all off-shell contributions into Bethe-Salpeter kernels.
Each diagram in the expansion contains 
\(\widetilde \sigma = \widetilde \sigma(q',p')\) and 
\(\widetilde \sigma^\dagger = \widetilde \sigma^\dagger(q,p)\) 
``endcaps'' on the far left and far right
respectively. The form of these functions depends on the original
interpolating fields, and does not affect the final answer.

Between endcaps each diagram contains some number of two-to-two and
three-to-three Bethe-Salpeter kernels. The two-to-two Bethe-Salpeter
kernel \(i B_{2 \rightarrow 2}\) is the sum of all diagrams which
are two-particle irreducible in the s-channel.
Given our constraint $m<E^*<5m$, together with the fact that
$B_{2\to2}$ always appears alongside a spectator line,
it follows that none of the $B_{2\to2}$ in 
Fig.~\ref{fig:fullskelexpansion} 
can have on-shell intermediate states. This implies that the integrands
``inside'' $B_{2\to2}$ are non-singular, and momentum sums can
be replaced by integrals. Thus we can replace the finite-volume
version of $B_{2\to2}$ with its infinite-volume correspondent.
Similarly \(i B_{3 \rightarrow 3}\) is defined so that it contains no diagram
in which three propagators carry the total energy-momentum 
\((E, \vec P)\). Diagrams with one propagator carrying \((E,\vec P)\), as
well as any odd number greater than three, are allowed. Again we drop
exponentially suppressed corrections and work with the infinite-volume
version of the kernel.

Finally, in our skeleton expansion all kernels and interpolating
functions are connected by fully-dressed propagators
\begin{equation}
\label{eq:propdef}
\Delta(q) \equiv \int d^4 x e^{i q x} \langle 0 \vert \mathrm{T}
\phi(x) \phi(0) \vert 0 \rangle\,.
\end{equation}
Here \(\phi(x)\) is a one particle interpolating field defined with
on-shell renormalization so that
\begin{equation}
\lim_{q^0 \rightarrow \omega_q} \Big [ \Delta (q) \, (q^2-m^2)/i \Big
] = 1 \,.
\end{equation}
We use infinite-volume fully-dressed propagators throughout, which is
justified because the self-energy graphs do not contain on-shell
intermediate states.

In summary, only three-particle intermediate states give important
(power-law) finite-volume corrections. 
The skeleton expansion
therefore keeps these on-shell states explicit and groups all off-shell
states into infinite-volume kernels and propagators.

\section{3-body phase space and notation}
\label{sec:notation}

To avoid repetitive definitions, we describe here the
coordinates and notation we use for three particles states,
and also collect some general notation.
We begin with the latter:
\begin{equation}
\int_{\vec k} \equiv \int d^3k/(2\pi)^3\,,
\ \ \
\omega_k = \sqrt{\vec k^2+m^2}\,,
\ \ \
\omega_{ka} = \sqrt{(\vec P-\vec k-\vec a)^2 + m^2}
\,.
\end{equation}
For three on-shell particles having total energy-momentum
$(E,\vec P)$, we parametrize phase space by the
momentum of one particle (the ``{spectator}''), 
usually denoted $\vec k$ or $\vec p$,
and the direction of one of the other two particles
in their CM frame, usually $\hat q^*$ or $\hat a^*$. 
Note that the magnitude of the momentum in the two-particle
CM frame is fixed by the {spectator} momentum. If the latter
is $\vec k$, this magnitude is denoted by $q_k$, and given by
\begin{equation}
\label{eq:qkxk}
\frac{q_k^2}{m^2} = x_k -1\,,\quad
x_k =  \frac{(E-\omega_k)^2 - (\vec P-\vec k)^2}{4 m^2}\,.
\end{equation}
Thus $x_k=1$ at the two-particle threshold.

In finite volume, three-particle phase space is restricted.
We parametrize it by a {spectator} momentum which now takes
on quantized values, $\vec k \in (2 \pi/L)  \mathbb Z^3$,
together with the angular momentum in the CM-frame of the
other two particles, $\ell, m$.
Thus three-to-three scattering amplitudes
become matrices with indices $k,\ell,m$, where $k$ is
a shorthand for all the quantized values of $\vec k$.
This is the natural extension of the two-particle analysis
in which the matrix indices are simply $\ell, m$.

Finally, we need notation for the case where two of the
three particles are on-shell, with four-momenta, say, $(\omega_p, \vec p)$
and $(\omega_k, \vec k)$, but the third, 
with four-momentum $(E - \omega_p - \omega_k, \vec P - \vec p - \vec k)$, 
is not.
Treating $\vec k$ as the {spectator}, we can boost to the 
zero-momentum frame of the other two particles, using a boost 
with velocity $\vec \beta_k=-(\vec P-\vec k)/(E-\omega_k)$.
This is possible kinematically  as long as $x_k>0$. 
We then denote $(\omega_{p^*},\vec p^*)$ as the
four-vector obtained by such a boost acting on $(\omega_p,\vec p)$,
with $p^*$ and $\hat p^*$ being the magnitude and
direction of $\vec p^*$. 
Since we allow the \((\vec P - \vec p - \vec k)\)-particle to be off-shell, 
\(p^*\) is not constrained. Similar definitions hold with $\vec k$ and $\vec p$
interchanged.

\section{Cusp effects and pole prescriptions}
\label{sec:cusp}

Our method for picking  out finite-volume corrections
is a generalization of that used in Ref.~\cite{Kim:2005} to analyze
the two-particle case. To illustrate the method, consider the
simplest contribution to the skeleton expansion,
the first diagram in Fig.~\ref{fig:fullskelexpansion}. 
Labeling its contribution \(\mathcal V_L\), 
and that of the corresponding infinite-volume diagram
\(\mathcal V_\infty\), one can show that
(up to exponentially small corrections)
\begin{equation}
\label{eq:noinsdiff}
\mathcal V_L - \mathcal V_\infty = \bigg [ \frac{1}{L^6} \sum_{\vec k,
    \vec a} - \int_{\vec k, \vec a} \bigg ] \frac{i \sigma(\vec k,
  \vec a) \sigma^\dagger(\vec k, \vec a)}{2 \omega_k 2 \omega_a 2
  \omega_{ka} (E - \omega_k - \omega _a - \omega_{ka} + i \epsilon)}
\,.
\end{equation}
Here $\vec k$ and $\vec a$ are the momenta flowing in the bottom
and top lines, respectively.
We have subtracted $\mathcal{V}_\infty$ because this 
leads to simpler expressions while not affecting the pole structure
(since $\mathcal{V}_\infty$ only has cuts).
In particular, the difference is dominated by 
the region of summation where the summand is singular, 
i.e. the region where all three-particles are on-shell. 
It is because of this singularity that the sum-integral
difference cannot be neglected.

To proceed, it turns out to be very useful to replace the integral
\(\int_k\) in the rightmost term in Eq.~(\ref{eq:noinsdiff}) with
the corresponding sum. This allows \(\mathcal V_L\) to be combined
 with the diagrams
on the second line of Fig.~\ref{fig:fullskelexpansion},
all of which have a ``spectator'' line carrying momentum $\vec k$.
This replacement is justified if 
\begin{equation}
\label{eq:kfunc}
  \int_{ \vec a} \frac{i \sigma(\vec k, \vec a) \sigma^\dagger(\vec k,
    \vec a)}{2 \omega_k 2 \omega_a 2 \omega_{ka} (E - \omega_k -
    \omega _a - \omega_{ka} + i \epsilon)}
\end{equation}
is a smooth function of \(\vec k\) with characteristic width \(m\). 
Unfortunately, although (\ref{eq:kfunc}) is finite and continuous, it is
not smooth due to a unitary cusp. This cusp is simply the imaginary part from the physical cut,
which turns on with infinite derivative
when one moves from below to above the two-particle threshold. The threshold occurs when 
$x_k=1$ [see Eq.~(\ref{eq:qkxk}) above].
This cusp problem arises because we have a third particle whose
momentum we are varying---it is not present in the two-particle analysis.

We can remove the cusp by subtracting, instead of $\mathcal{V}_\infty$,
a different infinite-volume quantity, $\widetilde{\mathcal{V}_\infty}$,
in which the pole prescription is changed.
We are free to do this as long as $\widetilde{\mathcal{V}_\infty}$
does not contain poles, as is the case here.
Specifically, we change from the \(i \epsilon\) prescription to 
a modified principal-value prescription, similar to that introduced in
Ref.~\cite{Polejaeva:2012}. It is defined by
\begin{equation}
\widetilde{\mathrm{PV}} \frac{1}{E - \omega_a - \omega_k -
  \omega_{ka}} \equiv \big [1 - i H(\vec k) H(\vec a) \,\mathrm{Im}
  \big] \frac{1}{E - \omega_a - \omega_k - \omega_{ka} + i \epsilon}
\,.
\label{eq:PVtilde}
\end{equation}
Here \(H(\vec k)\) is a smooth cutoff function, 
whose role is to damp contributions from subthreshold momenta. 
From analyzing more complicated diagrams, we find that it
is convenient to require%
\footnote{An example which satisfies all requirements is 
$H(\vec k ) \equiv J(x_k)$ with
\begin{equation}
J(x) = 0 \mathrm{\ for\ } x<0,\ \ \ \ J(x) = \exp \left( -(1/x) \exp
\left [-1/(1-x) \right] \right ) \mathrm{\ for\ } 0<x<1,
\ \ \ \ J(x)=1 \mathrm{\ for\ } 1<x \,.
\end{equation}}
\begin{equation}
\label{eq:HvalB}
H(\vec k)  = 1 \ \mathrm{if}\ x_k>1 \ \ \mathrm{and}\ \
H(\vec k)  = 0 \ \mathrm{if}\ x_k < 0
\,,
\end{equation}
\begin{equation}
\label{eq:Hsmooth}
\bigg [\frac{1}{L^3} \sum_{\vec k} - \int_{\vec k} \bigg] H(\vec k ) =
\mathcal O(e^{- m L}) \,.
\end{equation}
In the region where \(H(\vec k)H(\vec a)=1\), 
corresponding to all three particles being above threshold, 
our prescription becomes  ``\(1 - i \,\mathrm{Im} = \mathrm{Re}\)''.
This is the standard principal value prescription. 
It removes the unitary cusp, provided that we correctly analytically 
continue below threshold. More specifically, for subthreshold momenta, 
i.e.~those for which which \(E - \omega_k - \omega_a - \omega_{ka}<0\), 
no pole prescription is needed, so one might think that \(H\) should be
required to vanish. However, such a choice produces a cusp
which leads to power-law finite-volume corrections. We therefore define
\(\widetilde {\mathrm{PV}}\) below threshold by analytic continuation
of the above-threshold result, which avoids cusps. 
This is implicitly included in the definition of ``Im'' in 
Eq.~(\ref{eq:PVtilde}).
The product \(H(\vec k) H(\vec a)\) then provides a smooth
interpolation to the naive subthreshold definition. 

Putting everything together we obtain, for the first diagram,
\begin{equation}
\mathcal V_L - \widetilde{\mathcal V_\infty} = \frac{1}{L^3}
\sum_{\vec k} \bigg [ \frac{1}{L^3} \sum_{ \vec a} -
  \widetilde{\mathrm{PV}} \int_{ \vec a} \bigg ] \frac{i \sigma(\vec
  k, \vec a) \sigma^\dagger(\vec k, \vec a)}{2 \omega_k 2 \omega_a 2
  \omega_{ka} (E - \omega_k - \omega _a - \omega_{ka} )} \,.
\end{equation}
This is now in a form which allows us to use a generalization of
the result of Ref.~\cite{Kim:2005} in which ``sum minus integral''
acting on a pole picks out the residue of the pole (which is
an on-shell quantity) multiplied by a kinematic function
(related to L\"uscher's zeta-function~\cite{Luescher:1986n2,Luescher:1991n1}).
The main changes from Ref.~\cite{Kim:2005} are that the kinematic
function, called $\widetilde F$ below, differs slightly due to the
change in pole prescription, and that on-shell quantities which are analytically
continued below threshold appear. 

It is straightforward to extend this analysis to the diagrams
on the second line of Figure \ref{fig:fullskelexpansion}. These diagrams have any number of
two-to-two Bethe-Salpeter kernels appearing on the 
same pair of propagators. We denote the sum of all such diagrams, together with \(\mathcal V_L\), by
\(C_L^{(1)}\) and label the corresponding infinite-volume quantity
\(C_\infty^{(1)}\). We find that
\begin{equation}
\label{eq:C1res}
C_L^{(1)} - C_\infty^{(1)} = (\sigma + A'^{(1)}) 
\frac{i \widetilde F}{2 \omega   L^3} 
\frac{1}{1 + \widetilde{ \mathcal K }\; \widetilde F } 
(\sigma^\dagger + A^{(1)}) 
- (2/3) \sigma \frac{i \widetilde F}{2 \omega L^3} \sigma^\dagger \,.
\end{equation}
This expression has the form 
(row vector)\(\times\)(matrix)\(\times\)(column vector) in the direct
product space
\begin{equation}
\label{eq:matspace}
[\mathrm{finite\ volume\ spectator\ momentum\ } \vec k \in 
(2 \pi/L)  \mathbb Z^3]
\times [\mathrm{angular\ momentum\ }(\ell,m)] \,,
\end{equation}
which was introduced in Sec.~\ref{sec:notation}.
This means that the rows (\(\sigma\), \(A'^{(1)}\)) and columns
(\(\sigma^\dagger\), \(A^{(1)}\)) have one set of indices \(k, \ell,
m\) where \(k\) is short for \(\vec k \in (2 \pi/L) \mathbb Z^3\) and
where \(\ell, m\) describe the angular momentum in the CM-frame
of the non-spectator pair. Here $A'^{(1)}$ and $A^{(1)}$ are corrections
to the endcaps $\sigma$ and $\sigma^\dagger$, respectively, involving
insertions of $B_{2\to2}$. Their detailed form is irrelevant for the
finite-volume spectrum.
The matrices \([2 \omega   L^3]^{-1}\), \(\widetilde F\) and
\(\widetilde {\mathcal K}\) have two sets of
\(k, \ell, m\) indices. They are defined as
\begin{align}
\label{eq:omegamatdef}
\left[ \frac{1}{2 \omega L^3} \right]_{k',\ell',m';k,\ell,m} & \equiv
\delta_{k',k} \delta_{\ell',\ell} \delta_{m',m} \frac{1}{2 \omega_k
  L^3} \,, \ \ \ \ \ \ \ \ \ \ \widetilde{\mathcal
  K}_{k',\ell',m';k,\ell,m} \equiv \delta_{k',k} \,
\widetilde{\mathcal K}_{2;\ell',m';\ell,m}(\vec k) \,,\\
\label{eq:Fdef1}
\widetilde{F}_{k', \ell',m';k,\ell,m} & \equiv \delta_{k,k'} \frac{1}{2}
 \left[\frac{1}{L^3} \sum_{\vec a} - \widetilde{\mathrm{PV}}\!\!
   \int_{\vec a} \right] \frac{ {4 \pi} Y_{\ell',m'}(\hat a^*)
   Y_{\ell,m}^*(\hat a^*) H(\vec k) H(\vec a)}{2 \omega_a 2
   \omega_{ka}(E - \omega_k - \omega_a - \omega_{ka})}
 (a^*/q_k^*)^{\ell+\ell'} \,.
\end{align}
Here $\widetilde{\mathcal{K}}_2$ is the two-particle K-matrix defined
with our modified PV prescription. Its argument $\vec k$
indicates the momentum carried by the spectator, a notation
used frequently below. The quantities
appearing in Eq.~(\ref{eq:Fdef1}) are defined in Sec.~\ref{sec:notation}.

Aside from the spectator,
the diagrams leading to the result (\ref{eq:C1res})
are exactly those contributing to the two-particle
correlator studied in Ref.~\cite{Kim:2005}. 
Thus we expect the results to be closely related. 
Indeed, the matrix appearing in the first term on the r.h.s.~of 
(\ref{eq:C1res}) has the same form as the result of Ref.~\cite{Kim:2005}.
The second term is present here because of a mismatch
of symmetry factors. If, however, one considers a theory with
two identical particles plus a third which is
non-interacting (therefore non-identical), then the second term
is absent. Indeed, in this alternative theory we have already summed all
possible diagrams, and the divergence of the first term determines the
finite-volume spectrum. 

This observation provides a check on the formalism presented so far.
In the alternative theory we know the spectrum to be that of
a single free particle (with any finite-volume momentum, $\vec k$)
combined with that of two interacting particles in the box with 
combined momentum $\vec P-\vec k$. The latter spectrum is itself known
from Refs.~\cite{Rummukainen:1995vs,Kim:2005}, and
is obtained from the solutions of
\begin{equation}
\label{eq:KSSQC2}
\det\left[\mathcal{M}_2^{-1}(\vec k) + F_2(\vec k)\right] = 0\,.
\end{equation}
Here $i\mathcal{M}_2$ is the two-particle scattering amplitude,
while $F_2$ is a finite-volume kinematic factor.
Both are matrices in $\ell, m$ space.
$F_2$ is defined like $\widetilde F$
[Eq.~(\ref{eq:Fdef1})] except that the $k$ index
and $\delta_{k,k'}$ term are absent,
and the pole is regulated using the $i\epsilon$ prescription.
As above, the argument $\vec k$ indicates the spectator momentum,
so that the total momentum flowing through the two-particle correlator
is $(E-\omega_k,\vec P-\vec k)$.

Our expression for all diagrams with one particle unscattered 
[Eq.~(\ref{eq:C1res}) minus the last term]
gives the spectrum for a free spectator combined with that obtained
from the solutions of
\begin{equation}
\label{eq:ourQC2}
\det\left[\widetilde{\mathcal{K}}_2^{-1}(\vec k) 
+ \widetilde F_2(\vec k)\right] = 0\,.
\end{equation}
Here $\widetilde F_2$ is same as $\widetilde F$, except that
the $k$ index and $\delta_{k,k'}$ are removed.
These two results for the theory with one free particle 
agree, because Eqs.~(\ref{eq:KSSQC2})
and (\ref{eq:ourQC2}) turn out to be equivalent. This is because
\begin{equation}
\label{eq:KtoM2}
\mathcal{M}_2^{-1}(\vec k) -\widetilde{\mathcal{K}}_2^{-1}(\vec k) 
=\widetilde F_2(\vec k) - F_2(\vec k) =
-i \frac{q^*_k}{32 \pi } [q^{*2}_k + m^2 ]^{-1/2} H(\vec k) \mathbf{1}\,.
\end{equation}
Here all quantities are matrices in $\ell,m$ space, with 
the r.h.s.~proportional to the identity. 
Above the two-particle threshold, where $H(\vec k)=1$, 
the r.h.s.~is simply the imaginary part of \(\mathcal{M}_2^{-1}\),
so that $\widetilde{\mathcal{K}}_2^{-1}$ is the real part,
which is one way of defining the standard K-matrix.

\section{Three-particle singularities}

We next turn our attention to the finite-volume diagrams on the third
line of Figure \ref{fig:fullskelexpansion}, and in particular
the parts of the diagrams within the dashed rectangles. We will not state the
result of summing all such diagrams here. Instead we only comment that
the summation contains a factor of the form
\begin{equation}
i \widetilde{\mathcal K}_{3 \rightarrow 3}^{\ 2, \mathrm{unsym}}(\vec
k, \hat a'^*; \vec p, \hat a^*) \equiv 
i \widetilde {\mathcal K}_2(\vec k) 
[\widetilde{\mathrm{PV}} \Delta(P-p-k)] 
i \widetilde {\mathcal  K}_2(\vec p) \,.
\end{equation}
Here the superscript \(2\) on the l.h.s.~indicates that the quantity 
has two \(\widetilde{\mathcal K}_2\) insertions, while ``unsym'' indicates that it 
is not symmetric under the exchange of external momenta.
This is our first contribution to three-to-three scattering.
We have used \(\vec p, \hat a^*\) to parametrize three
on-shell particles, as explained in Sec.~\ref{sec:notation}.
Now observe that 
\(\widetilde{\mathcal K}_{3 \rightarrow 3}^{\ 2, \mathrm{unsym}}\) is singular at
\((P-p-k)^2=m^2\). As a result, the full 
\(\widetilde{\mathcal K}_{3   \rightarrow 3}\), 
defined as the sum of all connected three-to-three
diagrams with \(\widetilde{\mathrm{PV}}\) pole prescription, is
singular and does not have a uniformly convergent partial wave
expansion. It is therefore difficult to imagine how 
\(\widetilde {\mathcal K}_{3 \rightarrow 3}\) can be
directly extracted from the finite-volume spectrum. 

Indeed, it turns out the the quantization condition depends directly
not on \(\widetilde{\mathcal K}_{3 \rightarrow 3}\) but instead on a
subtracted quantity which is everywhere smooth. Furthermore, the terms
that we subtract depend only on on-shell two-to-two K-matrices, which
is reasonable since the divergences are due to on-shell intermediate
states. This means that one can recover 
\(\widetilde{\mathcal K}_{3 \rightarrow 3}\) from the spectrum, 
by first finding the divergence-free object and then adding in the 
known singular terms. The divergence-free three-to-three
K-matrix is defined as
\begin{equation}
\label{eq:Mdfdef}
i \Kdf {}(\vec k', \hat a'^{*}, \vec k, \hat a^{*}) \equiv i
\widetilde {\mathcal K}_{3 \rightarrow 3}(\vec k', \hat a'^{*}, \vec
k, \hat a^{*}) - \mathcal S \Big[i \mathcal D(\vec k', \hat a'^{*},
  \vec k, \hat a^*)\Big] \,,
\end{equation}
where \(\mathcal S\) denotes symmetrization of external momenta and
$\mathcal{D}$ satisfies an integral equation:
\begin{equation}
i \mathcal D(\vec k', \hat a'^{*}, \vec k, \hat a^*) = 4 \pi
Y_{\ell',m'}^*(\hat a'^{*}) i \mathcal D_{\ell', m'; \ell, m}(\vec k',
\vec k \,) Y_{\ell,m}(\hat a^*) \,,
\end{equation}
\begin{equation}
\label{eq:Ddef}
i \mathcal D(\vec k', \vec k) = i \widetilde {\mathcal K}_2(\vec k') i
G_{\infty}(\vec k', \vec k) i \widetilde {\mathcal K}_2(\vec k) +
\widetilde{\mathrm{PV}} \int_{\vec \ell} \frac{1}{2 \omega_{\ell}} i
\widetilde {\mathcal K}_2(\vec k') i G_\infty(\vec k', \vec \ell \,) i
\mathcal D(\vec \ell, \vec k) \,,
\end{equation}
Here $G_\infty$ is the three-particle pole transformed to two-particle
angular-momentum space, with additional factors included for
technical reasons:
\begin{equation}
\label{eq:Ginfdef}
 G_{\infty; \ell', m';\ell,m} (\vec k, \vec p \,) \equiv \frac{ 4 \pi
   Y_{\ell',m'}(\hat p^*) Y_{\ell,m}^*(\hat k^*) H(\vec k) H(\vec
   p\,)}{2 \omega_{pk} (E - \omega_p - \omega_k - \omega_{pk} )}
 (p^*/q_k^*)^{\ell'} (k^*/q_p^*)^{\ell} \,. 
\end{equation}
In Eq.~(\ref{eq:Ddef}), \(\mathcal D\), \( \widetilde{ \mathcal K}_2\)
and \( G_{\infty}\) are understood to have two sets of implicit
angular momentum indices, with internal indices contracted.
The starred quantities are defined in Sec.~\ref{sec:notation}.

Solving the integral equation (\ref{eq:Ddef}) iteratively,
the result, in schematic form, is 
$\mathcal{D}=\widetilde{\mathcal{K}}_2 G_\infty \widetilde{\mathcal{K}}_2+
\widetilde{\mathcal{K}}_2 G_\infty \widetilde{\mathcal{K}}_2 G_\infty \widetilde{\mathcal{K}}_2 +\dots$.
Thus the subtraction in (\ref{eq:Mdfdef}) removes the infinite
sequence of possibly divergent contributions to
$\widetilde {\mathcal K}_{3 \rightarrow 3}$. More precisely, 
the terms which are divergent are those for which the series of 
pairwise scatterings is possible classically. 
In the case of degenerate particles only the first two terms diverge. 
However, in a generalization to non-degenerate particles, 
the number of divergent diagrams will depend on the mass ratios. 
For this reason it is not surprising that our finite-volume analysis 
leads to a quantity with all terms subtracted. 

Since \(\Kdf {}\) is finite for all on-shell
momenta, it can be decomposed into spherical harmonics
\begin{equation}
\label{eq:Mdfdecom}
\Kdf {} (\vec k', \hat a'^{*}, \vec k, \hat a^{*}) 
\equiv 
4 \pi Y^*_{\ell', m'}(\hat a'^*) \Kdf{;k', \ell', m';k, \ell, m}
Y_{\ell,m}(\hat a^*) 
\,.
\end{equation}
Although \(\Kdf {;k', \ell', m';k, \ell, m}\) is defined for all real
\(\vec k\) and \(\vec p\), it turns out that our final answer will
only depend on the quantity at values \(\vec k, \vec p \in (2 \pi/L)
\mathbb Z^3\). The index notation on the left-hand side of
Eq.~(\ref{eq:Mdfdecom}) is meant to be suggestive of this
discretization. 
Indeed, from this point on \(\Kdf {;k', \ell', m';k,  \ell, m}\) 
is understood as a discrete matrix in the direct-product
space (\ref{eq:matspace}).  

\section{Three-particle quantization condition}

After a lengthy analysis, which we do not describe here,
we find the quantization condition\footnote{%
In the talks, we quoted a final result in which we used the
$i\epsilon$ prescription for infinite volume quantities rather
than our $\widetilde{\rm PV}$ prescription. The form of
the result was the same as Eq.~(\ref{eq:mainres}), but had 
${\mathcal{M}}_{{\rm df},3\to3}$ 
(the divergence-free scattering amplitude)
in place of 
$\widetilde{\mathcal{K}}_{{\rm df},3\to3}$,
and the definition of $\widetilde{F}_3$ was also slightly changed.
This result ignored the impact of cusp singularities 
and was incorrect (since not all power-law volume effects
were included). 
}
\begin{equation}
\label{eq:mainres}
\det \! \big [1 + \widetilde{F}_{3} \Kdf {} \big ] = 0 \,.
\end{equation}
The determinant is over the direct product space (\ref{eq:matspace}). 
At fixed \(\{L, \vec P\}\) the determinant is a function of \(E\). 
The set of
solutions \(E_1, E_2, \cdots\) to Eq.~(\ref{eq:mainres}) give the
spectrum of the finite-volume theory.

The quantity $\widetilde{F}_3$ appearing in Eq.~(\ref{eq:mainres}) 
depends on the two-to-two K-matrix 
[packaged into the matrix $\widetilde{\mathcal{K}}$ 
defined in Eq.~(\ref{eq:omegamatdef})],
as well as the kinematic ``sum minus integral'' function $\widetilde{F}$ 
[Eq.~(\ref{eq:Fdef1})], and one new kinematic function $G$.
Our result is
\begin{equation}
\label{eq:F3def}
\widetilde{F}_{3} \equiv \frac{\widetilde{F}}{2 \omega L^3} \left[ 
\frac{1}{1 + [1 + \widetilde{\mathcal K}  G ]^{-1} 
                   \widetilde{\mathcal K}   \widetilde{F}}
- \frac23 \right] 
\,,
\end{equation}
All quantities are matrices with pairs of implicit $k,l,m$ indices.
The function $G$ is simply $G_\infty$ [Eq.~(\ref{eq:Ginfdef})]
evaluated for discrete momenta, aside from a simple change in overall
normalization:
\begin{align}
\label{eq:Gdef}
 G_{k, \ell', m' ; p, \ell, m} & \equiv \frac{1}{2 \omega_p L^3}
 \frac{ 4 \pi Y_{\ell',m'}(\hat p^*) Y_{\ell,m}^*(\hat k^*) H(\vec k)
   H(\vec p\,)}{2 \omega_{pk} (E - \omega_p - \omega_k - \omega_{pk})}
 (p^*/q_k^*)^{\ell'} (k^*/q_p^*)^\ell \,.
\end{align}

We now give a brief discussion of the result (\ref{eq:mainres}).
We first note the superficial similarity to the two-particle quantization
condition, already given in two forms above
[Eqs.~(\ref{eq:KSSQC2}) and (\ref{eq:ourQC2})],
which can also be written
\begin{equation}
\det \! \big[1 + \widetilde{F}_2 \widetilde{\mathcal{K}}_{2} \big ]=0\,.
\label{eq:QC2}
\end{equation}
This form provides a clear separation between infinite-volume
quantities related to scattering (here the two-to-two K-matrix)
and finite-volume effects entering through the kinematical
function $\widetilde{F}_2$. It shows that the spectrum depends
only on the infinite-volume scattering amplitude
[related to the K-matrix through Eq.~(\ref{eq:KtoM2})].

The same comments hold for our three-particle result,
Eq.~(\ref{eq:mainres}), though with some subtleties.
First, $\widetilde{F}_3$ is not simply a kinematical function,
but rather a kinematical function 
($\widetilde{F}$---which is simply $\widetilde{F}_2$ multiplied 
by $\delta_{k,k'}$)
decorated by contributions from two-to-two scattering.
One can see from Eq.~(\ref{eq:F3def}) that
this decoration consists of two nested geometric
series, one involving the two-particle K-matrix, and
the other the ``switch-factor'' $G$.
These arise, respectively, from diagrams in the second
and third/fourth lines of Fig.~\ref{fig:fullskelexpansion}.
Such ``decoration'' seems an unavoidable consequence of
moving from two to three particles.

The second subtlety concerns the appearance of sub-threshold
two-to-two amplitudes. 
This point was first noted and discussed in Ref.~\cite{Polejaeva:2012}.
They arise in $\widetilde{\mathcal{K}}$ when the spectator
momentum is such that the remaining two particles are below
threshold. Such amplitudes can be defined by analytically continuing
the two-particle K-matrix from the physical region, as is in fact
routinely done when using the two-particle quantization condition
to discuss bound states. The regulator function $H$ in our
pole prescription ensures that we only need to analytically continue
for a distance $\sim m$.

The final subtlety concerns the connection with infinite-volume
quantities. Our result shows that the spectrum is determined
by infinite-volume quantities related to scattering, as was
previously found in the non-relativistic treatment of
Ref.~\cite{Polejaeva:2012}.
However, the quantities which appear are not the scattering amplitudes
themselves, but differ in two ways. 
First, our result contains divergence-free three-particle objects.
Second, these are defined using the $\widetilde{\rm PV}$,
rather than the $i\epsilon$, pole prescription.
Thus they are a version of K-matrices having some regulator dependence
entering through the choice of the function $H$.
We do not think that either difference is a serious concern.
The full amplitude can be reconstructed from the divergence-free one
using the integral equation (\ref{eq:Ddef}).
Similarly, we can write an integral equation relating 
$\Kdf{}$ to ${\mathcal{M}}_{{\rm df},3\to3}$.
These equations can be solved numerically once the approximations
to the scattering amplitudes discussed below are made.
Thus, our intermediate quantity $\Kdf{}$ is directly related to
$\mathcal{M}_{3\to3}$.
Furthermore, it is arguable that $\Kdf{}$ is a more 
natural quantity to appear than $\mathcal{M}_{3\to3}$,
since it is a smooth, real function
(without cusps or singularities) and thus is simpler to approximate.

\section{Truncation}

Our quantization condition (\ref{eq:mainres}) is a formal equation
involving infinite-dimensional matrices.
The same is true of
the two-particle quantization condition (\ref{eq:QC2})
[or equivalently Eq.~(\ref{eq:KSSQC2})].
To make these equations useful in practice one must develop systematic
approximation schemes which truncate the matrices down to finite dimensions.

In the two-particle case, this is justified because of
the rapid decrease of elastic scattering amplitudes with increasing $\ell$ 
at fixed energy. Mathematically, one must assume that 
$\mathcal{M}_{2}$ {\em vanishes} for $\ell>\ell_{\rm max}$
(which implies that $\widetilde{\mathcal{K}}_2$ vanishes also). 
Then, even though $\widetilde{F}_2$ is not diagonal, 
one can show that the determinant truncates to
that of a $(2\ell_{\rm max}+1)^2$ matrix (with further simplifications due to
residual symmetries whose details depend on $\vec P$). The simplest case is
$\ell_{\rm max}=0$, and this is often the form used in practice.
In any case, the key point is that truncation leads to an algebraic
equation involving $\widetilde{\mathcal{K}}_{2}$ evaluated at the CM energy
$E^*=\sqrt{E^2-\vec P^2}$ (with $E$ the energy of the spectral line).

What happens in the $3\to3$ case? Here we have a larger index space,
involving the spectator momentum $\vec k$. The key observation, however,
is that the sum over $\vec k$ is self-truncating. 
To understand this, note that the four-momentum flowing into
the non-spectator pair is $p_2=(E-\omega_k,\vec P-\vec k)$. 
For fixed $(E,\vec P)$, as $|\vec k|$ increases, $p_2^2$ decreases below
the threshold value $4m^2$.
In this subthreshold region (for the non-spectator pair) the intermediate
3-particle state cannot go on shell, and the sum-integral
difference $\widetilde{F}$
[Eq.~(\ref{eq:Fdef1})] becomes exponentially suppressed.
Although formally true immediately below threshold,
$p_2^2$ must be reduced by
$\sim m^2$ below the threshold value $4m^2$, before the
exponential suppression becomes numerically significant.
This is because, roughly speaking, the exponent is $L (4 m^2-p_2^2)/m$.
This is why one must allow a range of subthreshold momenta to have
an accurate quantization condition.

Our definition of $\widetilde{F}$ includes, in addition, the regulator
function $H(\vec k)$, which smoothly sets $\widetilde{F}$ to zero as
$p_2^2$ approaches $0$ (which is well into the regime where
$\widetilde{F}$ is exponentially suppressed). 
Thus we have an exact truncation, at the cost
of an exponentially small error.

The upshot is that the $\vec k$ sum is truncated, with say $N$ terms to be
kept. One then assumes that $\widetilde{\mathcal{K}}_{2}$ can be truncated at
$\ell_{\rm max,2}$, as in the
two-particle case, {\em and} that $\Kdf{}$ can be
truncated at $\ell_{\rm max,3}$. Here the prior removal of the singular part 
of $\Kdf{}$ is key, since otherwise truncation is not justified.
Calling $\ell_{\rm max}$ the larger of $\ell_{\rm max,2}$ 
and $\ell_{\rm max,3}$,
one can then show that the determinant equation collapses to one for
matrices of dimension $[N(2\ell_{max}+1)]^2$. 

The end result is an algebraic equation involving the various non-vanishing
partial waves of $\widetilde{\mathcal{K}}_{2}$ and 
$\Kdf{}$,
the latter evaluated at the CM energy $E^*$ of the spectral line. 
Assuming that ${\mathcal{M}}_{2}$
is known from studies of two-particle spectra using the L\"uscher method,
and interpolating as necessary,
one gains information about $\Kdf{}$ from each spectral line.
To make progress, one would likely need to parametrize $\Kdf{}$
as a function of $\vec k$, $\ell$ and $m$, and then use as many spectral
lines as parameters to determine the latter
(e.g. by varying $\vec P$, but keeping $E^*$ fixed).

\section{Threshold expansion}

An important check on our formalism is provided by the threshold
expansion. This applies at $\vec P=0$,
where the lightest non-interacting
3-particle state has $E=3m$ (with each particle at rest).
Interactions shift this by $\delta E\propto 1/L^3$.
For large $L$, $\delta E\ll m$, and the particles are non-relativistic.
In addition, our analysis shows that the relativistic transitions
($3\to1$, $3\to5$, etc.) do not play an essential role, leading
to non-analyticities only at energies beyond the range we consider.
Thus one can calculate $\delta E$ in this limit 
using a non-relativistic theory. 
Such a calculation has been carried out in Refs.~\cite{Beane:2007,Tan:2007},
with $\delta E$ obtained through\footnote{%
Subsequent work has pushed this to one higher order, but we will not
need this result.}
${\cal O}(1/L^6)$.
Our formalism should reproduce this result.

Close to threshold one can truncate the two-particle interactions
to be purely s-wave ($\ell_{\rm max,2}=0$) and take 
$\Kdf{}$ to be a constant 
(implying $\ell_{\rm max,3}=0$ too).\footnote{%
At threshold $\Kdf{}$ and $\mathcal{M}_{\mathrm{df},3\to3}$ are equal.
Thus the results in this section are unaffected by the presence
of cusps. It turns out that they enter the threshold expansion
first at ${\cal O}(1/L^{10})$.}
Thus $\ell_{\rm max}=0$ and the only matrix structure is in
spectator-momentum space.
Furthermore, one can show that the dominant term in
$[\widetilde{F}_3]_{\vec k,\vec k'}$ is that with $\vec k=\vec k'=0$, 
other matrix elements being suppressed by powers of $1/L$. 
Indeed we will see that, to the order we are working, 
only the $\vec k=\vec k'=0$ entry needs to be included. 
This means that the quantization condition (\ref{eq:mainres}) collapses to
\begin{equation}
[\widetilde{F}_3]_{\vec 0,\vec 0}\, \Kdf{} = -1
\label{eq:F3threshold}
\,.
\end{equation}

It is now useful to rewrite $\widetilde{F}_3$ as
\begin{equation}
\widetilde{F}_3 =  \frac{\widetilde{F}}{2\omega L^3}
\left[\frac13 + 
\frac{1}{\widetilde{\mathcal{K}}_{2}^{-1} + \widetilde{F} + G}  
\widetilde{F}\right]\,.
\label{eq:F3v2}
\end{equation}
From (\ref{eq:F3threshold}) we know that $[\widetilde{F}_3]_{\vec 0,\vec 0}$
scales as $L^0$ (since $\Kdf{}$ is a constant).
However, $\widetilde{F}_3$ contains an explicit factor of $L^{-3}$.
One can show that $\widetilde{F}\sim L^0$ near threshold, so the only way
that the overall $L^{-3}$ factor can be canceled is if
$(\widetilde{\mathcal{K}}_{2}^{-1} + \widetilde{F} + G) \sim L^{-3}$.
Thus the $L^0$, $L^{-1}$ and $L^{-2}$ parts of this combination
must be canceled by tuning $\delta E$.
This turns out to determine the $L^{-3}$, $L^{-4}$ and
$L^{-5}$ parts of $\delta E$, respectively.
The $L^{-6}$ part is then determined by enforcing
Eq.~(\ref{eq:F3threshold}).
We stress that the full matrix inverse of 
$(\widetilde{\mathcal{K}}_{2}^{-1} + \widetilde{F} + G)$ must
be evaluated (to the order we are working), 
even though the external indices are fixed to $\vec 0$. 

After a moderately lengthy calculation we find
\begin{eqnarray}
\delta E &=& \frac{12\pi a}{mL^3}\left[
1 - \frac{a}{\pi L}\mathit{I}
+ \left(\frac{a}{\pi L}\right)^2(\mathit{I}^2+\mathit{J})
+ \left(\frac{a}{\pi L}\right)^3
(-\mathit{I}^3+\mathit{I}\mathit{J} + 15 \mathit{K}
- 16 \mathit{Q} - 8 \mathit{R})\right]
\nonumber\\
&&+ \frac{72 a^3\pi^2 r}{mL^6}
+ \frac{36 a^2\pi^2}{m^3 L^6}
- \frac{\Kdf{}}{48 m^3 L^6}
+ {\cal O}(L^{-7})
\,. \label{eq:deltaE}
\end{eqnarray}
Here $a$ is the 2-particle scattering length (with the nuclear physics
sign convention: positive for repulsion), and $r$ is the 
corresponding effective range. $\mathit{I}$, $\mathit{J}$ and $\mathit{K}$
are familiar finite-volume zeta functions
involving sums over integer vectors:
\begin{equation}
\mathit{I} 
= \sum_{\vec n\ne0}^{|\vec n|=\Lambda} \frac1{\vec n^2}-4\pi\Lambda\,,
\ \
\mathit{J} 
= \sum_{\vec n\ne0}\frac1{(\vec n^2)^2}\,, 
\ \
\mathit{K} 
= \sum_{\vec n\ne0}\frac1{(\vec n^2)^3}\,.
\end{equation}
They have known numerical values.
$\mathit{Q}$ and $\mathit{R}$ are more complicated quantities:
\begin{eqnarray}
\mathit{Q} = - 2048 \pi^6 L^3 m^3 
\!\!\!\!\sum_{\vec k\ne 0,\vec p\ne 0}\!\!\!\!
G_{\vec 0,\vec k} G_{\vec k,\vec p} G_{\vec p,\vec 0}
&=& \!\!\sum_{\vec n_k\ne 0,\vec n_p\ne 0}^{\rm reg}
\frac1{\vec n_k^2\vec n_p^2 [\vec n_k^2+\vec n_p^2
+ (\vec n_k+\vec n_p)^2]}
+ {\cal O}(L^{-1})
\label{eq:Q}
\\
\mathit{R} = - 4096 \pi^6 L^3 m^3 
\!\sum_{\vec k\ne 0}\!
G_{\vec 0,\vec k} \widetilde{F}_{\vec k,\vec k} G_{\vec k,\vec 0}
&=&\sum_{\vec n_k\ne 0} \frac1{(\vec n_k^2)^2}
\sum_{\vec n_p}^{\rm reg}
\frac1{\vec n_k^2 + \vec n_p^2 + (\vec n_k+\vec n_p)^2}
+ {\cal O}(L^{-1})
\label{eq:R}
\end{eqnarray}
where $\vec k=2\pi \vec n_k/L$ and $\vec p=2\pi \vec n_p/L$.
For both $Q$ and $R$, the first form of the result shows how the matrix
nature (in spectator momentum space) 
of $G$ and $\widetilde{F}$ enters. In addition, both of these forms
are convergent sums. When expanded in the non-relativistic limit,
however, one ends up with divergent sums (the second expressions)
which must be regulated. It is the latter forms that appear in the
results of Refs.~\cite{Beane:2007,Tan:2007}.

The first line of our result for $\delta E$, Eq.~(\ref{eq:deltaE}),
agrees with the results of Refs.~\cite{Beane:2007,Tan:2007}.
Various aspects of this agreement are noteworthy.
The leading and first sub-leading terms ($L^{-3}$ and $L^{-4}$,
respectively) are simply the result for two-particles multiplied by
a factor of 3. This corresponds to the presence of
three possible two-particle pairs.
At ${\cal O}(L^{-5})$, however, features enter which are special
to the 3-particle case (leading to a flip in the sign of the 
$\mathit{J}$ term).
In addition, the switch factors $G$ play an essential role in our
calculation, and without their presence
we would not find agreement with Refs.~\cite{Beane:2007,Tan:2007}.

We turn now to the remaining $L^{-6}$ terms on the second line of
(\ref{eq:deltaE}). Here the situation is more complicated.
In the first of these terms, we have a factor of 72, while 
Ref.~\cite{Beane:2007} finds 24 and Ref.~\cite{Tan:2007} has 36.
The second of our terms (that proportional to $a^2$) is absent
in Refs.~\cite{Beane:2007,Tan:2007}.
And in the third of the terms, our K-matrix $\Kdf{}$
is replaced by an unphysical, regulator-dependent quantity.
For example, Ref.~\cite{Beane:2007} has $\eta_3(\mu)/L^6$,
with $\eta_3$ a contact 3-particle potential and $\mu$ a renormalization
scale.
Finally, Refs.~\cite{Beane:2007,Tan:2007} each have an additional
regulator-dependent term, e.g.
\begin{equation}
\frac{64\pi a^4}{mL^6} (3\sqrt3-4\pi)\log(\mu L)
\end{equation}
in Ref.~\cite{Beane:2007}. 
This term arises when dimensional regularization is used to 
define the second forms in Eqs.~(\ref{eq:Q},\ref{eq:R}).

We stress that, despite the discrepancies in the form of $L^{-6}$
terms, there is no conflict between our result and those of
Refs.~\cite{Beane:2007,Tan:2007}. We do not, {\em a priori}, know the
relation between $\Kdf{}$ and $\eta_3(\mu)$. One is a physical
quantity, while the other is a short-distance unphysical parameter,
as evidenced by its regulator dependence. It follows that the terms on the
second line of (\ref{eq:deltaE}) do not provide a consistency check,
but instead imply a relation between three-body parameters that is
required for the energy shifts to match. As far as we can see, there
is nothing forbidding this relation to include the finite $a^2$ and
$a^2 r$ terms. Indeed, a similar finite difference is required to
match  the results of Refs.~\cite{Beane:2007} and \cite{Tan:2007}.
Nevertheless, it would clearly be good to check this purported
relation in another context.

\section{Outlook}

Having a formalism is only the first step. We are presently investigating
the practicality of the (truncated) quantization condition using
simple models for the scattering amplitudes.
It is also important to understand more clearly the relationship of
our approach to those of Refs.~\cite{Polejaeva:2012,Briceno:2012rv} 
as well as to
that of HAL-QCD~\cite{HALQCD}.

We also plan to extend the theoretical work in several directions:
working with non-identical non-degenerate particles,
generalizing the Lellouch-L\"uscher relation \cite{Lellouch:2000pv} to three particles,
and extending our result to cases in which
the two-particle K-matrices can have poles above threshold,
i.e. in which there are two-particle resonances within the
kinematic range of our formula.
Ultimately, we aim to consider the four-particle case.

\section{Acknowledgments}
We thank Ra\'ul Brice\~no, Zohreh Davoudi and Akaki Rusetsky for
helpful discussions. MTH was supported in part
by a Fermilab Fellowship in Theoretical Physics.
Fermilab is operated by Fermi Research Alliance, LLC, under Contract
No.~DE-AC02-07CH11359 with the United States Department of Energy (DOE).
MTH and SRS were supported in part by DOE grant No.~DE-FG02-96ER40956.

\bibliographystyle{apsrev4-1}
\bibliography{ref}

\end{document}

%% file: fig25.pdf_t
\begin{picture}(0,0)%
\includegraphics{fig25.pdf}%
\end{picture}%
\setlength{\unitlength}{3947sp}%
\begingroup\makeatletter\ifx\SetFigFont\undefined%
\gdef\SetFigFont#1#2#3#4#5{%
  \reset@font\fontsize{#1}{#2pt}%
  \fontfamily{#3}\fontseries{#4}\fontshape{#5}%
  \selectfont}%
\fi\endgroup%
\begin{picture}(7938,3071)(2347,-3492)
\put(7772,-1912){\makebox(0,0)[lb]{\smash{{\SetFigFont{12}{14.4}{\rmdefault}{\mddefault}{\updefault}{\color[rgb]{0,0,0}$+$}%
}}}}
\put(3202,-1308){\makebox(0,0)[lb]{\smash{{\SetFigFont{12}{14.4}{\rmdefault}{\mddefault}{\updefault}{\color[rgb]{0,0,0}$+$}%
}}}}
\put(5251,-1912){\makebox(0,0)[lb]{\smash{{\SetFigFont{12}{14.4}{\rmdefault}{\mddefault}{\updefault}{\color[rgb]{0,0,0}$+$}%
}}}}
\put(2362,-699){\makebox(0,0)[lb]{\smash{{\SetFigFont{12}{14.4}{\rmdefault}{\mddefault}{\updefault}{\color[rgb]{0,0,0}$C_L(E, \vec P) =$}%
}}}}
\put(6042,-682){\makebox(0,0)[lb]{\smash{{\SetFigFont{12}{14.4}{\rmdefault}{\mddefault}{\updefault}{\color[rgb]{0,0,0}$+$}%
}}}}
\put(4326,-682){\makebox(0,0)[lb]{\smash{{\SetFigFont{12}{14.4}{\rmdefault}{\mddefault}{\updefault}{\color[rgb]{0,0,0}$+$}%
}}}}
\put(4778,-1308){\makebox(0,0)[lb]{\smash{{\SetFigFont{12}{14.4}{\rmdefault}{\mddefault}{\updefault}{\color[rgb]{0,0,0}$+$}%
}}}}
\put(6832,-1308){\makebox(0,0)[lb]{\smash{{\SetFigFont{12}{14.4}{\rmdefault}{\mddefault}{\updefault}{\color[rgb]{0,0,0}$+$}%
}}}}
\put(3197,-1912){\makebox(0,0)[lb]{\smash{{\SetFigFont{12}{14.4}{\rmdefault}{\mddefault}{\updefault}{\color[rgb]{0,0,0}$+$}%
}}}}
\put(3203,-2452){\makebox(0,0)[lb]{\smash{{\SetFigFont{12}{14.4}{\rmdefault}{\mddefault}{\updefault}{\color[rgb]{0,0,0}$+$}%
}}}}
\put(5741,-2447){\makebox(0,0)[lb]{\smash{{\SetFigFont{12}{14.4}{\rmdefault}{\mddefault}{\updefault}{\color[rgb]{0,0,0}$+$}%
}}}}
\put(3203,-2888){\makebox(0,0)[lb]{\smash{{\SetFigFont{12}{14.4}{\rmdefault}{\mddefault}{\updefault}{\color[rgb]{0,0,0}$+ \cdots$}%
}}}}
\put(3202,-3323){\makebox(0,0)[lb]{\smash{{\SetFigFont{12}{14.4}{\rmdefault}{\mddefault}{\updefault}{\color[rgb]{0,0,0}$+$}%
}}}}
\put(5437,-3323){\makebox(0,0)[lb]{\smash{{\SetFigFont{12}{14.4}{\rmdefault}{\mddefault}{\updefault}{\color[rgb]{0,0,0}$+$}%
}}}}
\put(7675,-3323){\makebox(0,0)[lb]{\smash{{\SetFigFont{12}{14.4}{\rmdefault}{\mddefault}{\updefault}{\color[rgb]{0,0,0}$+ \cdots$}%
}}}}
\put(8372,-670){\makebox(0,0)[lb]{\smash{{\SetFigFont{12}{14.4}{\rmdefault}{\mddefault}{\updefault}{\color[rgb]{0,0,0}$+ \cdots$}%
}}}}
\put(8759,-2444){\makebox(0,0)[lb]{\smash{{\SetFigFont{12}{14.4}{\rmdefault}{\mddefault}{\updefault}{\color[rgb]{0,0,0}$+ \cdots$}%
}}}}
\put(9334,-1302){\makebox(0,0)[lb]{\smash{{\SetFigFont{12}{14.4}{\rmdefault}{\mddefault}{\updefault}{\color[rgb]{0,0,0}$+ \cdots$}%
}}}}
\end{picture}%